\documentclass[12pt]{article} 
\PassOptionsToPackage{hyphens}{url}\usepackage{hyperref}
\pdfoutput=1

\usepackage{longtable}
\usepackage{cclicenses}
\usepackage[official]{eurosym}
\usepackage{amsbsy}
\usepackage{upgreek}
\renewcommand{\vec}[1]{\mathbf{#1}}

\title{Growth, degrowth, and the challenge of artificial superintelligence\footnote{Accepted manuscript. Journal reference: \textit{Pueyo, S., 2018. Growth, degrowth, and the challenge of artificial superintelligence. Journal of Cleaner Production 197, 1731-1736} (\url{https://doi.org/10.1016/j.jclepro.2016.12.138}). This work is distributed under a Creative Commons 4.0 License, CC-BY-NC-ND. \cc \byncnd}}

\author{Salvador Pueyo\textsuperscript{a,b,}\thanks{E-mail: spueyo@riseup.net}\\\textit{\small{\textsuperscript{a}Dept. Evolutionary Biology, Ecology, and Environmental Sciences, }}\\\textit{\small{Universitat de Barcelona, Av. Diagonal 645, 08028 Barcelona,}}\\\textit{\small{Catalonia, Spain}}\\\textit{\small{\textsuperscript{b}Research \& Degrowth, C/ Trafalgar 8 3, 08010 Barcelona, Catalonia, Spain}}} 
\date{} 
\begin{document} 
\maketitle 
\begin{abstract} 
\noindent The implications of technological innovation for sustainability are becoming increasingly complex with information technology moving machines from being mere tools for production or objects of consumption to playing a role in economic decision making. This emerging role will acquire overwhelming importance if, as a growing body of literature suggests, artificial intelligence is underway to outperform human intelligence in most of its dimensions, thus becoming \textit{superintelligence}. Hitherto, the risks posed by this technology have been framed as a technical rather than a political challenge. With the help of a thought experiment, this paper explores the environmental and social implications of superintelligence emerging in an economy shaped by neoliberal policies. It is argued that such policies exacerbate the risk of extremely adverse impacts. The experiment also serves to highlight some serious flaws in the pursuit of economic efficiency and growth per se, and suggests that the challenge of superintelligence cannot be separated from the other major environmental and social challenges, demanding a fundamental transformation along the lines of degrowth. Crucially, with machines outperforming them in their functions, there is little reason to expect economic elites to be exempt from the threats that superintelligence would pose in a neoliberal context, which opens a door to overcoming vested interests that stand in the way of social change toward sustainability and equity.
\end{abstract}
\section{Introduction}
We could be approaching a technological breakthrough with unparalleled impact on the lives of every reader of this paper, and on the whole biosphere. It might seem fanciful to suggest that, in a near future, artificial intelligence (AI) could vastly outperform human intelligence in most or all of its dimensions, thus becoming \textit{superintelligence}. However, in the last few years, this position has been endorsed by a number of recognized scholars and key actors of the AI industry. Several research institutions have been created to explore the implications of superintelligence, for example at Oxford and Cambridge Universities. For details on how this idea emerged and is becoming established, see the chronological table in the Supplementary Material, and for a thorough understanding of the current discussions see Bostrom (2014) or Shanahan (2015).

\textit{Artificial intelligence} (AI) is defined as \textit{computational procedures for automated sensing, learning, reasoning, and decision making} (AAAI, 2009, p.\ 1). AIs can be programmed to pursue some given goals. For example, AIs programmed to win chess matches have been defeating human world champions since 1997 (Bostrom, 2014). Current AIs have narrow scopes, while a hypothetical superintelligence would be more effective than humans in pursuing virtually every goal. AI experts surveyed in 2012/13 assigned a probability of 0.1 to crossing the threshold of human-level intelligence by 2022, 0.5 by 2040 and 0.9 by 2075 (median estimates; M{\"u}ller et al., 2016). The European Commission recently launched the \euro{}1 billion Human Brain Project with the intent of simulating a complete human brain as early as 2023, but its chances of success have been questioned (Nature Editors, 2015), and superintelligence is thought to be more easily attainable by engineering it from first principles than by emulating brains (Bostrom, 2014).

Following Yudkowsky (2001), the current discussion on the implications of superintelligence (Bostrom, 2014; Shanahan, 2015) is framed around two possibilities: the first superintelligences to emerge will be either \textit{hostile} or \textit{friendly} (depending on their programmed goals). In most authors' views, these would result in either the worst or the best imaginable consequences for humanity, respectively\footnote{The techno-utopia of a world ruled by friendly superintelligence reveals extreme \textit{technological enthusiasm} and \textit{technocracy}, in Kerschner and Ehlers' (2016) terminology. Technocracy is also apparent in moves to avoid public implication in this issue (Supplementary Material).}. Much subtler distinctions apply to weaker forms of AI, but it is argued that intermediate outcomes are less likely for an innovation as radical as superintelligence (Bostrom, 2014, p.\ 20).

Hostile superintelligence is imagined as a result of failure to specify and program the desired goals properly, or of instability in the programmed goals, or less frequently as the creation of some illicit group. Therefore, it is framed as a technical rather than a political challenge. Most of the research is focused on ways to align the goals of a hypothetical superintelligence with the goals of its programmer (Sotala and Yampolskiy, 2015), without questioning the economic and political system in which AI is being developed. Kurzweil (2005, p.\ 420) is explicit in that an \textit{open free-market system} maximizes the likelihood of aligning AI with human interests, and is leading a confluence of major corporations to advance an agenda of radical techno-social transformation based on this and other allied technologies (Supplementary Material). The benefits imagined from friendly superintelligence find an economic expression in rates of growth at an order of magnitude above the traditional ones or more (Hanson, 2001, 2008; Bostrom, 2014).

This view is akin to that of some authors within sustainability science, who take seriously the environmental challenges posed by economic growth, technological innovation and the functioning of capitalist markets, but seek solutions based on these same elements. Opposed to this position is the idea of degrowth (D'Alisa et al., 2014). Degrowth advocates hold a diversity of views on technology (see the Introduction to this special issue), but agree that indefinite growth is not possible if measured in biophysical terms, and is not always desirable if measured as GDP, both for environmental and for social reasons. Also, they are critical of capitalist schemes: to foster a better life in a downsized economy, they would rather support redistribution, sharing, democracy and the promotion of non-materialistic and prosocial values.

The challenges of sustainability and of superintelligence are not independent. The changing fluxes of energy, matter, and information can be interpreted as different faces of a general acceleration\footnote{The perception of general technological and social acceleration is shared by authors close to degrowth (Rosa and Scheuerman, 2009) and by those concerned with superintelligence. The latter often suggest that acceleration will culminate in a singularity, related to the emergence of this form of AI (Supplementary Material).}. More directly, it is argued below that superintelligence would deeply affect production technologies and also economic decisions, and could in turn be affected by the socioeconomic and ecological context in which it develops. Along the lines of Pueyo (2014, Sec.\ 5), this paper presents an approach that integrates these topics. It employs insights from a variety of sources, such as ecological theory and several schools of economic theory.

The next section presents a thought experiment, in which superintelligence emerges after the technical aspects of goal alignment have been resolved, and this occurs specifically in a neoliberal scenario. Neoliberalism is a major force shaping current policies on a global level, which urges governments to assume as their main role the creation and support of capitalist markets, and to avoid interfering in their functioning (Mirowski, 2009). Neoliberal policies stand in sharp contrast to degrowth views: the first are largely rationalized as a way to enhance efficiency and production (Plehwe, 2009), and represent the maximum expression of capitalist values.

The thought experiment illustrates how superintelligence perfectly aligned with capitalist markets could have very undesirable consequences for humanity and the whole biosphere. It also suggests that there is little reason to expect that the wealthiest and most powerful people would be exempt from these consequences, which, as argued below, gives reason for hope. Section \ref{solucio} raises the possibility of a broad social consensus to respond to this challenge along the lines of degrowth, thus tackling major technological, environmental, and social problems simultaneously. The uncertainty involved in these scenarios is vast, but, if a non-negligible probability is assigned to these two futures, little room is left for either complacency or resignation.

\section{Thought experiment: Superintelligence in a neoliberal scenario}
\label{futur}

Neoliberalism is creating a very special breeding ground for superintelligence, because it strives to reduce the role of human agency in collective affairs. The neoliberal pioneer Friedrich Hayek argued that the \textit{spontaneous order} of markets was preferable over conscious plans, because markets, he thought, have more capacity than humans to process information (Mirowski, 2009). Neoliberal policies are actively transferring decisions to markets (Mirowski, 2009), while firms' automated decision systems become an integral part of the market's information processing machinery (Davenport and Harris, 2005). Neoliberal globalization is locking governments in the role of mere players competing in the global market (Swank, 2016). Furthermore, automated governance is a foundational tenet of neoliberal ideology (Plehwe, 2009, p.\ 23).

In the neoliberal scenario, most technological development can be expected to take place either in the context of firms or in support of firms\footnote{E.g., EU's Human Brain Project \textit{is committed to driving forward European industry} (HBP, n.d.).}. A number of institutionalist (Galbraith, 1985), post-Keynesian (Lavoie, 2014; and references therein) and evolutionary (Metcalfe, 2008) economists concur that, in capitalist markets, firms tend to maximize their growth rates (this principle is related but not identical to the neoclassical assumption that firms maximize profits; Lavoie, 2014). Growth maximization might be interpreted as expressing the goals of people in key positions, but, from an evolutionary perspective, it is thought to result from a mechanism akin to natural selection (Metcalfe, 2008). The first interpretation is insufficient if we accept that: (1) in big corporations, \textit{the managerial bureaucracy is a coherent social-psychological system with motives and preferences of its own} (Gordon, 1968, p.\ 639; for an insider view, see Nace, 2005, pp.\ 1-10), (2) this system is becoming \textit{techno-social-psychological} with the progressive incorporation of decision-making algorithms and the increasing opacity of such algorithms (Danaher, 2016), and (3) human mentality and goals are partly shaped by firms themselves (Galbraith, 1985).

The type of AI best suited to participate in firms' decisions in this context is described in a recent review in \textit{Science}: \textit{AI researchers aim to construct a synthetic homo economicus, the mythical perfectly rational agent of neoclassical economics. We review progress toward creating this new species of machine, }machina economicus (Parkes and Wellman, 2015, p.\ 267; a more orthodox denomination would be \textit{Machina oeconomica}).

Firm growth is thought to rely critically on retained earnings (Galbraith, 1985; Lavoie, 2014, p.\ 134-141). Therefore, economic selection can be generally expected to favor firms in which these are greater. The aggregate retained earnings\footnote{Here (like, e.g., in Lavoie, 2014), \textit{retained earnings} are the part of earnings that the firm retains, i.e., a flow. Other sources use \textit{retained earnings} to refer to the cumulative result of retaining earnings, i.e., a stock.} \textit{RE} of all firms in an economy can be expressed as:
\begin{equation}
\label{re}
RE=F_\vec{E} (\vec{R},\vec{L},\vec{K})-\vec{w} \cdot \vec{L}-(\vec{i}+\pmb{\updelta}) \cdot \vec{K}-g.		
\end{equation}

Bold symbols represent vectors (to indicate multidimensionality). $F$ is an aggregate production function, relying on inputs of various types of natural resources $\vec{R}$, labor $\vec{L}$ and capital $\vec{K}$ (including intelligent machines), and being affected by environmental factors\footnote{And also by technology and organization, but these are not introduced explicitly because they are assumed to affect every term of this equation. The inclusion of $\vec{R}$ and $\vec{E}$ and their multidimensionality rely on insights from ecological economics (e.g., Martinez-Alier, 2013).} $\vec{E}$; $\vec{w}$ are wages, $\vec{i}$ are returns to capital (dividends, interests) paid to households, $\pmb{\updelta}$ is depreciation and $g$ are the net taxes paid to governments.

Increases in retained earnings face constraints, such as trade-offs among different parameters of Eq.\ \ref{re}. The present thought experiment explores the consequences of economic selection in a scenario in which two sets of constraints are nearly absent: sociopolitical constraints on market dynamics are averted by a neoliberal institutional setting, while technical constraints are overcome by asymptotically advanced technology (with extreme AI allowing for extreme technological development also in other fields). The environmental and the social implications are discussed in turn. Note that this scenario is not defined by some contingent choice of AIs' goals by their programmers: The goals of maximizing each firm's growth and retained earnings are assumed to emerge from the collective dynamics of large sets of entities subject to capitalistic rules of interaction and, therefore, to economic selection.

\subsection{Environment and resources}
\label{ambient}

Extreme technology would allow maximizing $F$ in Eq.\ \ref{re} for some given $\vec{R}$ and $\vec{E}$, but would also alter the availability of resources $\vec{R}$ and the environment $\vec{E}$ indirectly. Would there still be relevant limits to growth? How would these transformations affect welfare?

To address the first question, let us consider growth in different dimensions:

\begin{itemize}
\item
Energetic throughput: It is often thought that the source that could allow \textit{energy production} (meaning tapping of exergy) to keep on increasing in the long term is nuclear fusion. This will depend on whether it is physically possible for controlled nuclear fusion to reach an energy return on energy investment EROI $>> 1$ (Hall, 2009). Even in this case, new limits would be eventually met, such as global warming due to the dissipated heat by-product (Berg et al., 2015). This same limit applies to other sources, such as space-based solar power. It is not known how global warming and other components of $\vec{E}$ would affect $F$ in a superintelligent economy, or the potential for mitigation or adaptation with a bearable energetic cost. Whatever the sources of energy eventually used, the constraints on growth are likely to become less stringent right after the development of superintelligence, but this bonus could be exhausted soon if there is a substantial acceleration of growth.
\item
Other components of biophysical throughput: Economies use a variety of resources with different functions, subject to their own limits. However, extreme technological knowledge would allow collapsing the various resource constraints into a single energetic constraint, so energy could become a common numeraire. The mineral resources that have been dispersed into the environment can be recovered at an energetic cost (Bardi, 2010). Currently, many constraints on biological resources cannot be overcome by spending energy (e.g., the overexploitation of some given species), but this will change if future developments in nanotechnology, genetic engineering or other technologies are used to obtain goods reproducing the properties that create market demand for such resources.
\item
Information processing: Information processing has a cost in terms of resources. Operating energy needs pose an obstacle to brain emulations with current computers (Sandberg, 2016), but the hardware requirements (Sandberg, 2016) could be met soon (Hsu, 2016), and other paths to superintelligence could be more efficient (Sandberg, 2016). However, current ICT relies on a variety of elements that are increasingly scarce (Ragnarsd{\'o}ttir, 2008). In principle, closing their cycles once they are dispersed in the environment has an enormous energetic cost (Bardi, 2010). The resource needs of future intelligent devices are unknown, but could limit their proliferation. This does not have to be incompatible with a continued increase in their capabilities: When ecosystems reach their own environmental limits, biological production stagnates or declines, but, often, there is a succession of species with increasing capacity to process information (Margalef, 1980).
\item
GDP: Potentially, it could continue to increase without need of growth in biophysical throughput, e.g., through trade in online services. It is argued in Sec. \ref{society} that this could well happen without benefiting human welfare.
\end{itemize}

Superintelligence holds the potential for extreme ecoefficiency: In the terms of Eq.\ \ref{re}, firms could not only increase $F$ given $\vec{R}$, but also decrease depreciation $\pmb{\updelta}$ (which, however, would only be viable for assets that do not need quick innovation because of competition). Increasing resource efficiency and decreasing turnover are common in maturing ecosystems (Margalef, 1980). However, ecoefficiency does not suffice to prevent impacts on the environment $\vec{E}$ (which does not only affect production but also the welfare of humans and other sentient beings). With firms maximizing their growth with few legal constraints (as corresponds to the type of society envisaged in Sec. 2.2), extreme resource efficiency could well entail an extreme rebound effect (Alcott, 2015), which is tantamount to generalized ecological disruption.

\subsection{Society}
\label{society}

The literature on superintelligence foresees enormous benefits if superintelligent devices are aligned with market interests, including tremendous profits for the owners of capital (Hanson, 2001, 2008; Bostrom, 2014). By simple extrapolation of shorter-term prognoses (Frey and Osborne, 2013; see also van Est and Kool, 2015), this literature also anticipates huge technological unemployment, but Bostrom (2014, p.\ 162) claims that, with an astronomic GDP, the trickle down of even minute amounts in relative terms would result in fortunes in absolute terms. However, if there were astronomic growth (e.g., focused on the virtual sphere) while food or other essential goods remained subject to environmental constraints and competition between basic needs and other uses, resulting in mounting prices, a minute income in relative terms would be minute in its practical usefulness, and most people might not benefit from this growth, or even survive (think, e.g., of the role of biofuels in recent famines; Eide, 2009). In fact, there are even more basic aspects of the standard view that are debatable. This section presents a different view, building on the assumption that firms generally tend to maximize growth under environmental constraints. The following points discuss the resulting changes in each of the social parameters in Eq. 1, and relate them to broader changes in society:

\begin{itemize}
\item
$\vec{L}$: A continuing trend toward $\vec{L}=\vec{0}$ is plausible, but it could also be reversed because of resource scarcity. Following Sec.\ \ref{ambient}, energetic cost could be the main factor to decide between humans or machines in functions that do not need large physical or mental capacities. Humans are made up of elements that follow relatively closed cycles and are easily available, while most current machines use nonrenewable materials whose availability is declining irreversibly (Georgescu-Roegen, 1971). Intelligent devices could thus become quite costly (Sec.\ \ref{ambient}). A variety of responses are imaginable, from finding techniques to build machines with more sustainable materials to creating machine-biological hybrids or modified humans; yet, it cannot be taken for granted that human work would be discarded. Initially, one extra reason to use human workers would be the big stock available. Even if human labor persisted, some major changes would be foreseeable: (1) Pervasive \textit{rationalization} maximizing the output extracted from labor inputs. Current experience with digital firms point to insidious techniques of labor management to the detriment of workers' interests (Mosco, 2016). (2) AIs replacing humans in important functions that need large mental capacities. These include the senior managers of big corporations and other key decision makers (as well as people devoted to economically relevant creative or intellectual tasks). A few \textit{unmanned} companies already exist (Cruz, 2014).
\item
$\vec{w}$: Thus far, $\vec{w}$ and $\vec{L}$ seem to have been affected similarly by IT, via labor demand (Autor and Dorn, 2013). However, it is worth noting that firms also have an impact on human wants (Galbraith, 1985), and that this impact is being enhanced by AI. Every user of the Internet is already interacting daily with forerunners of \textit{Machina oeconomica} that manage targeted advertising (Parkes and Wellman, 2015). \textit{Relational artifacts} (Turkle, 2006) promise an even more sophisticated manipulation of human emotions. There is empirical evidence that, as it would be expected, the compulsion to consume induced by advertising results in longer working hours and depressed wages (Molinari and Turino, 2015). Furthermore, consumption is not the only motivation to work (Weber, 1904); e.g., some firms induce workers to identify with them (Galbraith, 1985). If these trends continued to the extreme, humanity would become extremely addicted to consumption and to work, and wages would drop to the minimum needed to survive and work (assuming that human labor remains competitive; otherwise, $\vec{w}$ would be reduced to the zero vector $\vec{0}$).
\item
$\vec{i}$: Like work, having capital invested in firms is not just motivated by the wish to consume (Weber, 1904). Procedures like inducing identification (Galbraith, 1985) could magnify the other motivations and reduce $\vec{i}$. Consumption advertising acts in this case as a conflicting pressure (Molinari and Turino, 2015), but firms paying profits to households would probably be outcompeted by firms with no effective ownership (technically, nonprofits) or owned by other firms, which would allow reducing $\vec{i}$ to $\vec{0}$ (note that dividends and interests paid to other firms, including banks, cancel out because Eq. 1 refers to the aggregate of all firms). The owners of capital might currently have an economic function by allocating resources, but automated stock-trading systems have already determined between half and two thirds of U.S.\ equity trading in recent years (Karppi and Crawford, 2015), making human participation increasingly redundant.
\item
Demand: This is not an explicit term in Eq.\ \ref{re}, but is implicit in $F$ to the extent that production is addressed to the market. In an economy in which humans receive minimum wages and no profits, or in an economy without humans, demand would be basically reduced to firms' investment demand. This would serve no purpose, but would result from economic selection favoring firms with the greatest growth rate. Given the complex interactions mediated by demand, it is unclear whether or not a maximization of each firm's growth should translate to a maximization of aggregate growth.
\item
$g$: For a strict neoliberal program, the main role of governments would be to serve markets, and this function would determine some $g$ negotiated with firms. Directly or indirectly, governments would continue to exert functions of surveillance and coercion, aided by vast technological advances. Their decisions would be increasingly automated, whether or not they maintained some nominal power for human policy makers. Even elections are starting to be mediated by intelligent advertising (Mosco, 2016).
\end{itemize}

Therefore, a range of negative impacts can be expected, and they are unlikely to spare senior managers or capital owners.

Let us consider some moderate deviations from this political extreme. For example, these effectively \textit{selfish} automated firms could coordinate to address shared problems such as resource limitations, but this does not mean that they would seek to benefit society, such as by ceding resources for people's use with no benefit for firms' growth. Or, before superintelligence is fully developed, governments could try to implement some model combining market competition as a force of technological innovation and wealth creation with economic and technological regulations to ensure that humans (in general, or some privileged groups) obtain some share of the wealth that is produced. However, this project would meet some formidable obstacles:

\begin{enumerate}
\item
Ongoing neoliberal globalization is making it increasingly difficult to reverse the transfer of power to markets. A reversal will also be increasingly unlikely as computerization permeates and creates dependence in every sphere of life and the capacity of firms to shape human preferences increases.
\item
The mere prohibition of some features in AIs\footnote{This would be one of the few types of regulation that appear to be acceptable from a neoliberal viewpoint, taking Hayek (1966) as a reference.} poses technical problems that could prove intractable. In the words of Russell (interviewed by Bohannon, 2015): \textit{The regulation of nuclear weapons deals with objects and materials, whereas with AI it will be a bewildering variety of software that we cannot yet describe. I’m not aware of any large movement calling for regulation either inside or outside AI, because we don’t know how to write such regulation}.
\item
The objective role of humans obtaining profits from this type of firms would be parasitic. Parasites extract resources from organisms that surpass them in information and capacity of control (Margalef, 1980). In nature, parasites generally have high mortality rates, but persist by reproducing intensively. No equivalent strategy can be imagined in this case. The transfer of profits to humans would be an ecological anomaly, likely to be unstable in a competitive framework. 
\end{enumerate}

A much more likely departure from strict neoliberalism would result from structural mutations that would carry the system even further from any human plan, in unpredictable manners. Such mutations were excluded from the definition of this scenario, but not because they should be unlikely. In particular, they could provide a path to forms of \textit{hostile superintelligence} more similar to those in the literature.

Marxists believe that societies dominated by one social class can be the breeding ground for newer hegemonic social classes. In this way bourgeois would have displaced aristocrats, and they expect proletarians to displace the bourgeois (Marx and Engels, 1888). However, the bourgeoisie represented an advance in information processing and control, unlike the proletariat. AIs are better positioned to become hegemonic entities (even if unconsciously). This would not be just a social transition, but a biospheric transition comparable to the displacement of RNA by DNA as the main store of genetic information. So far, there is nothing locking future superintelligences in the service of human welfare (or the welfare of other sentient beings). Whether and how this future world would be shaped by the type of society from which it emerges is extremely uncertain, but neoliberalism can be seen as a blueprint for a Kafkaesque order in which humans are either absent or exploited for no purpose, and ecosystems deeply disturbed.

\section{Degrowth as a viable alternative}
\label{solucio}

Criticisms to the environmental and social impacts of the capitalist market are often answered with appeals to the gains in \textit{efficiency} and long-term growth brought about by a \textit{free} market. The above thought experiment shows how misleading it is to assume that efficiency and growth are intrinsically beneficial. The economic system as a whole may become larger and more efficient, but there is nothing in its \textit{spontaneous order} guaranteeing that the whole will serve the interest of its human parts. This becomes even more evident when approaching the point in which humans could cease to be the most intelligent of the elements interacting in this complex system. Even though the thought experiment assumes neoliberal policies, as one of the purest expressions of pro-capitalist policies, Sec.\ \ref{society} also lists some reasons to be skeptical of reformist solutions.

Here, a response to this challenge is outlined. This involves, first of all, to disseminate it and integrate it into a general criticism of the logic of growth and a search for systemic alternatives, in contrast to the \textit{technocratic} (sensu Kerschner and Ehlers, 2016) strategies to keep the management of this issue within limited circles (Supplementary Material). This awareness could initially permeate the social movements that originated in reaction to a variety of environmental and social problems caused by the current growth-oriented economy (including the incipient resistances to labor models introduced by digital firms; Mosco, 2016).

This will not just be one more addition to a list of dire warnings like resource exhaustion, environmental degradation and social injustice: While the economic elites now have the means to protect themselves from all of these threats, it is shown above that intelligent devices could well end up replacing them in their roles, thus equating their future to that of the rest of humanity. This alters the nature of the action for system change. It means that, in fact, this action does not oppose the interests of the most influential segments of society. A new role for social movements is to help these elites (and the rest of humanity) understand which policies are really in their best interest. In the kind of alternatives outlined below, such elites will gradually lose their privileges, but they will gain a much better life than if the loss of privileges occurs in the way that Sec.\ \ref{futur} suggests. Initially, few in the elites will be ready for such a radical change in their worldview, but these few could start a snowball effect. This is a game-changer creating new, previously  unimaginable opportunities.

A key step will be to reform the process of international integration. Rather than democracy controlled by the global market, markets will need to be democratically controlled (there has been a long-standing search for alternatives, e.g., The Group of Green Economists, 1992). This will not necessarily have to be followed by a trajectory toward a fully planned economy: a lot of research needs to be done on new ways to benefit from democratically \textit{tamed} self-organization processes (Pueyo, 2014). What does not suffice, however, is the old recipe of setting some minimum constraints with the expectation that, then, the forces of market competition will be harnessed for the general interest. If, as suggested in Sec.\ \ref{society}, there is no way for governments to control a mass of entities evolving in undesirable ways, an alternative is to deflect the forces that drive such evolution. This entails nothing less than moving from an economic system that promotes self-interest, competitiveness, and unlimited material ambitions in firms and individuals to a system that promotes altruism, collective responsibility, and sufficiency. In short, moving from the logic of growth to the logic of degrowth (see D'Alisa et al., 2014).

Thus, besides regulations setting constraints of various types, there is a need for methods to align economic selection with the collective interests. The application of such methods would, for example, cause demand (which affects production $F$ in Eq. 1) to become positively correlated with wages (i.e., with each firm's contribution to $\vec{w}$), negatively correlated with resource use ($\vec{R}$), and properly correlated with other more subtle parameters (not explicit in Eq.\ \ref{re}). The \textit{common good economy} (Felber, 2015) is an approach worth considering because it aims explicitly to remove pressures that propel growth, and is already expanding with the involvement of many businesses. In this approach, a key tool is the \textit{common good balance sheet}, a matrix of indicators of firms' social and environmental performance designed by participatory means, completed by the firms and (ideally) revised by independent auditors. Its function is to ease the application of ethical criteria by private and public agents interacting with firms in every stage of production and consumption. Felber (2015) envisions an advanced stage in which firms and the whole economy transcend their current nature (e.g., big firms would be democratized). While the common good balance sheet would serve mainly as an aid to change firms' general goals, it could also incorporate some explicit indicator of the perilousness of the software that these firms develop or use.

Hopefully, changing values in firms, governments, and social movements will also ease the change in individual values. This will further reduce the risk of having people engaged in the development of undesirable forms of AI. Furthermore, for those still engaged in such activities, there will be an increased chance of others in their social networks detecting and interfering with their endeavor. This reorientation at all levels (from the individual to the international sphere) will also help to address forms of AI distinct but no less problematic than \textit{Machina oeconomica}, such as autonomous weapons.

Even with such transformations, it will not be easy to decide democratically the best level of development of AI, but the types of AI should become less challenging. (Also, these transformations could moderate the pace of technological change and make it more manageable, by relaxing the competitive pressure to innovate). However, they will only be viable if they take place before reaching a possible point of no return, which could occur well before superintelligence emerges (considering irreversibility, obstacle 1 in Sec.\ \ref{society}).

\section{Conclusions}

There is little predictability to the consequences that superintelligence will have if it does emerge. However, the thought experiment in Sec.\ \ref{futur} suggests some special reasons for concern if this technology is to arise from an economy forged by neoliberal principles. While this experiment draws a disturbing future both environmentally and socially, it also opens the door to a much better future, in which not only the challenges of superintelligence but many other environmental and social problems are addressed. This pinch of optimism has two foundations: 1) The thought experiment suggests that nobody is immune to this threat, including the economically powerful, which makes it less likely that the action to address it gets stranded on a conflict of interests. 2) The neutralization of this threat could need systemic change altering the very motivations of economic action, which would ally the solution of this problem with the solution of many other obstacles to a sustainable and fair society, along the lines of degrowth. One of the main dangers now lies in our hubris, which makes it so difficult to conceive of anything ever defying human hegemony.

\section*{Acknowledgements}

I am grateful to Centre de Recerca Matem{\`a}tica (CRM) for its hospitality,  to Melf-Hinrich Ehlers for calling my attention on Mirowski and other useful comments, to Aaron Vansintjan for proofreading the manuscript and for useful comments, and, also for their useful comments, to Linda Nierling , Laura Blanco, Anna Palau, {\`A}lex Tortajada, S{\'i}lvia Heras and the anonymous reviewers.

\section*{References}

AAAI, 2009. Interim Report from the Panel Chairs. AAAI Presidential Panel on Long-Term AI Futures. Available at: \url{https://www.aaai.org/Organization/Panel/panel-note.pdf} (accessed 03-06-2015).\\

\noindent Alcott, B., 2014. Jevon's paradox (rebound effect), in: D’Alisa, G., Demaria, F., Kallis, G. (Eds.), 2014. Degrowth: A Vocabulary for a New Era. Routledge, London, pp.\ 121-124.\\

\noindent Autor, D. H., Dorn, D., 2013. The growth of low-skill service job and the polarization of the US labor market. Am.\ Econ.\ Rev.\ 103 , 1553-1597.\\

\noindent Bardi, U., 2010. Extracting minerals from seawater: an energy analysis. Sustainability 2, 980-992.\\

\noindent Berg, M., B. Hartley, Richters, O., 2015. A stock-flow consistent input-output model with applications to energy price shocks, interest rates, and heat emissions. New J.\ Phys.\ 17, 015011.\\ 

\noindent Bohannon, J., 2015. Fears of an AI pioneer. Science 349, 252.\\

\noindent Bostrom, N., 2014. Superintelligence: Paths, Dangers, Strategies. Oxford University Press.\\

\noindent Cruz, K., 2014. Exclusive interview with BitShares. Bitcoin Magazine, 8.10.2014. Available at: \url{https://bitcoinmagazine.com/16972/exclusive-interview-bitshares/}(accessed 30.08.2015).\\

\noindent D’Alisa, G., Demaria, F., Kallis, G. (Eds.), 2014. Degrowth: A Vocabulary for a New Era. Routledge, London. \\

\noindent Danaher, J., 2016. The threat of algocracy: Reality, resistance and accommodation. Philos.\ Technol., doi: 10.1007/s13347-015-0211-1.\\

\noindent Davenport, T.H., Harris, J.G., 2005. Automated decision making comes of age. MIT Sloan Manage. Rev. 46(4), 83-89.\\

\noindent Eide, A., 2009. The Right to Food and the Impact of Liquid Biofuels (Agrofuels). FAO, Rome.\\ 

\noindent Felber, C., 2015. Change Everything. Creating an Economy for the Common Good. Zed Books, London.\\

\noindent Frey, C.B., Osborne, M.A., 2013. The future of employment: How susceptible are jobs to computerisation? Oxford University. Available at: \url{http://www.oxfordmartin.ox.ac.uk/publications/view/1314} (accessed 03.06.2015). \\

\noindent Galbraith, J.K. 1985. The New Industrial State, 4th ed. Houghton Mifflin, Boston.\\

\noindent Georgescu-Roegen, N., 1971. The Entropy Law and the Economic Process. Harvard University Press, Cambridge, MA.\\ 

\noindent Gordon, S., 1968. The close of the Galbraithian system. J.\ Polit.\ Econ.\ 76, 635-644.\\

\noindent Hall, C.A.S., Balogh, S., Murphy, D.J.R., 2009. What is the minimum EROI that a sustainable society must have? Energies 2, 25-47.\\

\noindent Hanson, R.D., 2001. Economic growth given machine intelligence. Available at: \url{http://hanson.gmu.edu/aigrow.pdf} (accessed 09.08.2015).\\

\noindent Hanson, R.D., 2008. Economics of the singularity. IEEE Spectrum 45(6), 45-50.\\ 

\noindent Hayek, F.A., 1966. The principles of a liberal social order. Il Politico 31, 601-618.\\

\noindent HBP, n.d. Overview. Available at: \url{https://www.humanbrainproject.eu/2016-overview} (accessed 28.04.2016.).\\

\noindent Hsu, J, 2016. Power problems threaten to strangle exascale computing. IEEE Spectrum, 08.01.2016. Available at: \url{http://spectrum.ieee.org/computing/hardware/power-problems-threaten-to-strangle-exascale-computing} (accessed 17.04.2016.).\\

\noindent Karppi, T., Crawford, K. 2016. Social media, financial algorithms and the Hack Crash.\ Theor.\ Cult.\ Soc.\ 33, 73-92.\\

\noindent Kerschner, C., Ehlers , M.-H., 2016. A framework of attitudes towards technology in theory and practice. Ecol.\ Econ.\ 126, 139-151.\\

\noindent Kurzweil, R., 2005. The Singularity Is Near: When Humans Transcend Biology. Duckworth, London.\\

\noindent Lavoie, M., 2014. Post-Keynesian Economics: New Foundations. Edward Elgar, Cheltenham, UK.\\

\noindent Margalef, R., 1980. La Biosfera entre la Termodin{\'a}mica y el Juego. Omega, Barcelona.\\ 

\noindent Martinez-Alier, J., 2013. Ecological Economics, in: International Encyclopedia of the Social and Behavioral Sciences, Elsevier, Amsterdam, p.\ 91008.\\

\noindent Marx, K., Engels, F., 1888. Manifesto of the Communist Party (English version).\\

\noindent Metcalfe , J.S., 2008. Accounting for economic evolution: Fitness and the population method. J.\ Bioecon.\ 10, 23-49.\\

\noindent Mirowski, P., 2009. Postface, in: Mirowski, P., Plehwe, D. (Eds.), The Road from Mont P{\`e}lerin. Harvard University Press, pp.\ 417-455.\\

\noindent Molinari, B., Turino, F., 2015. Advertising and aggregate consumption: A Bayesian DSGE assessment. Working Papers (Universidad Pablo de Olavide, Dept. Economics) 15.02. Available at: \url{http://www.upo.es/econ/molinari/Doc/adv_rbc15.pdf}.\\

\noindent Mosco, V., 2016. Marx in the cloud, in: Fuchs, C., Mosco, V. (Eds.), Marx in the Age of Digital Capitalism. Brill, Leiden, pp.\ 516-535.\\

\noindent M{\"u}ller, V.C., Bostrom, N., 2016. Future progress in artificial intelligence: A survey of expert opinion, in: M{\"u}ller, V.C. (Ed.), Fundamental Issues of Artificial Intelligence. Springer, Berlin, pp.\ 553-571.\\

\noindent Nace, T., 2005. Gangs of America. Berrett-Koehler, San Francisco, CA.\\

\noindent Nature Editors, 2015. Rethinking the brain. Nature 519, 389.\\ 

\noindent Parkes, D. C., Wellman, M. P., 2015. Economic reasoning and artificial intelligence. Science 349, 267-272.\\

\noindent Plehwe, D., 2009. Introduction, in: Mirowski, P., Plehwe, D. (Eds.), The Road from Mont P{\`e}lerin. Harvard University Press, pp.\ 1-42.\\ 

\noindent Pueyo, S., 2014. Ecological econophysics for degrowth. Sustainability 6, 3431-3483. \url{https://ecoecophys.files.wordpress.com/2015/03/pueyo-2014.pdf}\\

\noindent Ragnarsd{\'o}ttir, K.V., 2008. Rare metals getting rarer. Nat.\ Geosci.\ 1, 720-721.\\ 

\noindent Rosa, H., Scheuerman, W.E., 2009. High-Speed Society. Pennsylvania State University Press.\\

\noindent Sandberg, A., 2016. Energetics of the brain and AI. Tech.\ Rep.\ STR 2016-2. Available at: \url{arXiv:1602.04019v1}\\

\noindent Shanahan, M., 2015. The Technological Singularity. MIT Press, Cambridge, MA.\\

\noindent Sotala, K., Yampolskiy, R.V., 2015. Responses to catastrophic AGI risk: A survey. Phys.\ Scripta 90, 018001.\\

\noindent Swank, D., 2016. Taxing choices: international competition, domestic institutions and the transformation of corporate tax policy. J.\ Eur.\ Public Policy 23, 571-603.\\

\noindent The Group of Green Economists, 1992. Ecological Economics: A Practical Programme for Global Reform. Zed Books, London.\\

\noindent Turkle, S., 2006. Artificial intelligence at 50: From building intelligence to nurturing sociabilities. Dartmouth Artifical Intelligence Conference, Hanover, NH, 15-07-2006. \url{http://www.mit.edu/~sturkle/ai@50.html}\\

\noindent van Est, R., Kool, L., 2015. Working on the Robot Society. Rathenau Instituut , The Hague.\\

\noindent Weber, M., 1904. Die protestantische Ethik und der ``Geist'' des Kapitalismus. Part 1. Archiv f{\"u}r Sozialwissenschaft und Sozialpolitik 20, 1-54.\\

\noindent Yudkowsky, E., 2001. Creating Friendly AI 1.0: The Analysis and Design of Benevolent Goal Architectures. The Singularity Institute, San Francisco, CA. Available at: \url{https://intelligence.org/files/CFAI.pdf}

\newpage
\setcounter{page}{1}
\setcounter{footnote}{0}

\section*{Supplementary Material}

\noindent \textbf{Table S1.} How the concept of superintelligence emerged and spread, and how the manegement of superintelligence is addressed. A chronology (up to August 2015).

\begin{longtable}{|ll|p{13cm}|}

\hline

& &\\

1863& &Just four years after the publication of \textit{On the Origin of Species}, Cellarius \cite{Cellarius1863} observed that, already in that times, the speed of technological evolution was incomparably quicker than the speed of biological evolution. He concluded that, if humanity did not leave the path of industrialization, at some point \textit{man will have become to the machine what the horse and the dog are to man}.\\

& &\\

\hline

& &\\

1950& &
Alan Turing \cite{Turing1950}, who set the foundations of modern computers, thought that human-level artificial intelligence would be reached relatively soon and theorized about it.\\

& &\\

\hline

& &\\

1958& &John von Neumann (another protagonist of the early stages of computers and AI) was reported  as noting \textit{the ever accelerating progress of technology and changes in the mode of human life, which gives the appearance of approaching some essential singularity in the history of the race beyond which human affairs, as we know them, could not continue} \cite{Ulam1958}. In this way von Neumann introduced the concept of \textit{singularity}, which is often identified with the point of time at which superintelligence emerges \cite{Sandberg2010}.\\

& &\\

\hline

& &\\

1961& &The beginner of cybernetics, Norbert Wiener \cite{Wiener1961}, warned of ways in which AI could have catastrophic outcomes.\\

& &\\

\hline

& &\\

1965& &The statistician I.J. Good \cite{Good1965} introduced the idea of an \textit{intelligence explosion}, in which an AI would enhance itself recursively thus becoming an \textit{ultraintelligent machine}.\\

& &\\

\hline

& &\\

1993& &In a keynote speech to NASA, Vernor Vinge \cite{Vinge1993} built on von Neumann and Good to predict a \textit{coming technological singularity} characterized by the onset of \textit{superhuman intelligence} or \textit{superintelligence}. He expressed a deep \textit{deterministic pessimism} (expression in \cite{Kerschner2016}) about this transition. In his speech, he described some discussions on this topic that were already going on, and put forward the essentials of the current debate.\\

& &\\

\hline

& &\\

2000& &The \textit{Singularity Institute for Artificial Intelligence} (SIAI, or just SI) was created in California \cite{SU2012}, its main goal being the development of what its researcher Eliezer Yudkowsky \cite{Yudkowsky2001} called a \textit{friendly} AI with human-equivalent or \textit{transhuman} mind\footnote{This is a strong instance of \textit{technocracy} (terminology in \cite{Kerschner2016}), as a one-sided attempt to develop a device expected to replace human governance. One alleged motivation is to crowd out a possible \textit{hostile} superintelligence. Technocratic strategies are frequent in this field, as apparent also from notes 3-6.}.\\

& &\\

\hline

& &\\

2005& &Professor Nick Bostrom founded the \textit{Future of Humanity Institute} of the University of Oxford \cite{FHI2015a}, whose flagship topic is the future impact of superintelligence.\\

& &\\

& &Publication of \textit{The Singularity is Near} \cite{Kurzweil2005}, the best known of a series of books in which Ray Kurzweil (who would be appointed Director of Engineering at Google in 2012 \cite{Kurzweil2015}) popularized a \textit{transhumanist}\footnote{\textit{Transhumanists} advocate the use of technology to radically alter the human condition in biological and other aspects \cite{Bostrom2005}. It is an instance of \textit{technological enthusiasm} and \textit{technocracy} (terminology in \cite{Kerschner2016}).} view of superintelligence and other technologies. It became a New York Times bestseller.\\

& &\\

\hline

& &\\

2006& &Ray Kurzweil and Peter Thiel (co-founder of Paypal) started the yearly \textit{Singularity Summit} under the auspices of the SI \cite{SU2012}. \\

& &\\

\hline

& &\\

2008& &The president of the Association for the Advancement of Artificial Intelligence (AAAI) brought together a group of leading AI researchers \textit{to explore and reflect about societal aspects of advances in machine intelligence}, in the \textit{AAAI 2008-09 Presidential Panel on Long-Term AI Futures}\footnote{This is another technocratic move (see note 1), considering the statement in \cite{AAAI2009b}, p.\ 62: \textit{We believe AAAI and AI researchers should take a leading role in dealing with the moral, ethical, and
legal issues involving AI systems (and not leave it to others!)}.}. Ref.\ \cite{AAAI2009b} (p.\ 20) suggests that the eventual attainment of human-level intelligence was taken for granted in this meeting. There was more skepticism about an \textit{intelligence explosion}, but some panelist recommended more research in this hypothesis \cite{AAAI2009}.\\

& &\\

& &Ray Kurzweil and Peter H. Diamandis started the \textit{Singularity University} \cite{SU2015a}, which can be interpreted as an attempt to accelerate the arrival of the singularity. Its corporate founders were Genenthec, Autodesk, Google, Cisco, Kauffman, Nokia and ePlanet Capital \cite{SU2015b}. This institution strives to expedite the deployment of what they call \textit{exponential technologies} (a constellation including AI but also other areas prioritized by transhumanists, such as biotechnology or nanotechnology) in corporations (from those created in their Startup Lab to established firms like Coca-Cola) and other contexts\footnote{Another technocratic move (see note 1).} \cite{SU2015a}.\\

& &\\

\hline

& &\\

2012& &The astrophysicist and former president of the Royal Society Martin Rees, the philosopher Huw Price and the co-founder of Skype Jaan Tallinn founded the Centre for the Study of Existential Risk (CSER) of the University of Cambridge, with the dangers of AI as its main topic \cite{CSER,Hui2012}.\\

& &\\

& &The Singularity University acquired the Singularity Summit from the SI, and also the SI brand \cite{SU2012}, with the SI becoming the \textit{Machine Intelligence Research Institute} MIRI \cite{MIRI2013}.\\

& &\\

\hline

& &\\

2014&April&Three well-known physicists (Stephen Hawking, Max Tegmark and the Nobel laureate Frank Wilczek) and Berkeley computer science professor and AI expert Stuart Russell wrote a journal paper \cite{Hawking2014} that attracted much attention to this issue. They stated: \textit{it's tempting to dismiss the notion of highly intelligent machines as mere science fiction. But this would be a mistake, and potentially our worst mistake ever}.\\

& &\\

&May&The \textit{Future of Life Institute} \cite{FLI2015b} was created, featuring among its founders and Scientific Advisory Board number of recognized scientists, AI engineers and CEOs of major corporations \cite{FLI2015c}. This institute declares to be \textit{currently focusing on potential risks from the development of human-level artificial intelligence} \cite{FLI2015c}.\\

& &\\

&July&Nick Bostrom (University of Oxford, Future of Humanity Institute) published \textit{Superintelligence: Paths, Dangers, Strategies} \cite{Bostrom2014}, widely perceived as the book of reference in this topic, and another New York Times Bestseller.\\

& &\\

&October&Public pronouncement on superintelligence by Elon Musk (co-founder and CEO of Tesla Motors). He declared: \textit{I think we should be very careful about artificial intelligence. If I had to guess at what our biggest existential threat is, it's probably that} \cite{McFarland2014}.\\

& &\\

\hline

& &\\

2015&January&Publication of the answers by 192 selected thinkers of Edge's \textit{annual question}: \textit{What do you think of machines that think?} \cite{Edge2015}.\\

& &\\

& &Pronouncement by Bill Gates \cite{Gates2015}: \textit{I am in the camp that is concerned about super intelligence. First the machines will do a lot of jobs for us and not be super intelligent. That should be positive if we manage it well. A few decades after that though the intelligence is strong enough to be a concern. I agree with Elon Musk and some others on this and don't understand why some people are not concerned}.\\

& &\\

& &Open Letter by a long list of recognized experts in AI under the auspices of the Future of Life Institute, stating that \textit{research on how to make AI systems robust and beneficial is both important and timely} \cite{FLI2015e}. Referring to systems that \textit{surpass human performance in most cognitive tasks}, the accompanying document on research priorities states: \textit{Assessments of this success probability vary widely between researchers, but few would argue with great confidence that the probability is negligible} \cite{FLI2015f}.\\

& &\\

& &Elon Musk donated \$10 million to the Future of Life Institute \textit{to run a global research program aimed at keeping AI beneficial to humanity}\footnote{Some interpret that \textit{Musk has decided to be vocal on the issue and to make this ``donation'' as a pre-emptive strike against negative public opinion, a potential obstacle for AI on its journey towards maturity and profitability} \cite{Mack2015}, which would entail another technocratic move (see note 1).} \cite{FLI2015a}.\\

& &\\

&July-August&While the present work is carried out, some symptoms of continuing interest in the topic are: (1) A new Open Letter released by the Future of Life Institute and signed by thousands of AI and robotics researchers, in this case to promote an international \textit{ban on offensive autonomous weapons}\footnote{The letter describes the horrors of these weapons but also admits that one of its motivations is to prevent a \textit{major public backlash against AI}, which is consistent with the technocratic moves mentioned in notes 1, 3, 4 and 5.} \cite{FLI2015g}. (2) A special issue in \textit{Science} on the \textit{Rise of the machines} \cite{Stajic2015}, featuring several reviews on AI and an interview to AI's pioneer Stuart Russell (CSER) warning on the dangers of superintelligences and other AIs \cite{Bohannon2015}. (3) The award of an ERC Advanced Grant to Nick Bostrom to pursue his research \cite{FHI2015b}. (4) MIT's publication of \textit{The Technological Singularity} by Murray Shanahan (Professor of Cognitive Robotics at Imperial College London) \cite{Shanahan2015}.\\

& &\\

\hline

\end{longtable}

\end{document}